\documentclass[11pt]{article}
\baselineskip
\pdfoutput=1
\usepackage{amsmath}
\usepackage{amssymb}
\usepackage{graphicx}
\usepackage{cite}
\usepackage{collref}
\usepackage{tabls}
\usepackage{hyperref}
\newcommand{\N}{\mathbb{N}}
\newcommand{\SU}{\mathrm{SU}}
\newcommand{\dd}{{\rm{d}}}
\newcommand{\Tr}{{\rm Tr\,}}

\newcommand{\Slat}{S_{\mbox{\tiny{lat}}}}

\newcommand{\Tc}{T_{\mbox{\tiny{c}}}}
\newcommand{\THagedorn}{T_{\mbox{\tiny{H}}}}
\newcommand{\npol}{n_{\mbox{\tiny{pol}}}}
\newcommand{\nbeta}{n_\beta}
\newcommand{\nconfig}{n_{\mbox{\tiny{conf}}}}
\newcommand{\dadj}{d_{\mbox{\tiny{a}}}}
\newcommand{\LambdaMSbar}{\Lambda_{\overline{\tiny\mbox{MS}}}}

\newcommand{\redchisq}{\chi^2_{\tiny\mbox{red}}}
\newcommand{\mthreshold}{m_{\mbox{\tiny{th}}}}

\begin{document}

\begin{titlepage}
\begin{center}
{\Large\bf Excluded-volume effects for a hadron gas in Yang-Mills theory}
\end{center}
\vskip1.3cm
\centerline{Paolo~Alba$^{a}$, Wanda~Maria~Alberico$^{b}$, Alessandro~Nada$^{b}$, Marco~Panero$^{b}$, and Horst~St\"ocker$^{a,c,d}$}
\vskip1.5cm
\centerline{\sl $^a$ Frankfurt Institute for Advanced Studies, Goethe Universit\"at Frankfurt}
\centerline{\sl Ruth-Moufang-Stra{\ss}e 1, D-60438 Frankfurt am Main, Germany}
\vskip0.5cm
\centerline{\sl $^b$ Department of Physics, University of Turin \& INFN, Turin}
\centerline{\sl Via Pietro Giuria 1, I-10125 Turin, Italy}
\vskip0.5cm
\centerline{\sl $^c$ Institut f\"ur Theoretische Physik, Goethe Universit\"at Frankfurt}
\centerline{\sl Max-von-Laue-Stra{\ss}e 1, D-60438 Frankfurt am Main, Germany}
\vskip0.5cm
\centerline{\sl $^d$ GSI Helmholtzzentrum f\"ur Schwerionenforschung GmbH}
\centerline{\sl Planckstra{\ss}e 1, D-64291 Darmstadt, Germany}
\vskip0.5cm
\begin{center}
{\sl  E-mail:} \hskip 1mm \href{mailto:alba@fias.uni-frankfurt.de}{{\tt alba@fias.uni-frankfurt.de}}, \href{mailto:alberico@to.infn.it}{{\tt alberico@to.infn.it}}, \href{mailto:anada@to.infn.it}{{\tt anada@to.infn.it}},\\ \href{mailto:marco.panero@unito.it}{{\tt marco.panero@unito.it}}, \href{mailto:stoecker@fias.uni-frankfurt.de}{{\tt stoecker@fias.uni-frankfurt.de}}
\end{center}
\vskip1.0cm
\begin{abstract}
When the multiplicities of particles produced in heavy-ion collisions are fitted to the hadron-resonance-gas model, excluded-volume effects play a significant r\^ole. In this work, we study the impact of such effects on the equation of state of pure Yang-Mills theory at low temperatures, comparing the predictions of the statistical model with lattice results. In particular, we present a detailed analysis of the $\SU(2)$ and $\SU(3)$ Yang-Mills theories: we find that, for both of them, the best fits to the equilibrium thermodynamic quantities are obtained when one assumes that the volume of different glueball states is inversely proportional to their mass. The implications of these findings for QCD are discussed.
\end{abstract}

\end{titlepage}

\section{Introduction}
\label{sec:intro}

The ongoing program of relativistic heavy-nuclei collisions at particle accelerators provides much experimental information about strongly interacting matter under extreme conditions of temperature and/or density~\cite{Andronic:2014zha, Heinz:2015tua, Akiba:2015jwa, Dainese:2016dea}. Most strikingly, it shows that, at temperatures $T \gtrsim 160$~MeV, a new state of matter exists, in which color charges are deconfined and chiral symmetry gets restored: the quark-gluon plasma (QGP)~\cite{Heinz:2000bk, Gyulassy:2004zy}.

This conclusion is derived from the concurrent observation of several, distinct phenomena, such as elliptic flow~\cite{ATLAS:2011ah, ATLAS:2012at, Chatrchyan:2012ta}, jet quenching~\cite{Aad:2010bu, Chatrchyan:2011sx, Chatrchyan:2012gt}, quarkonium suppression~\cite{Aad:2010aa, Chatrchyan:2011pe, Chatrchyan:2012np, Abelev:2012rv}, enhanced production of strange hadrons~\cite{Abelev:2006cs, Abelev:2007rw, Abelev:2008zk, Abelev:2013zaa}, as well as characteristic spectra of photons and leptons~\cite{Adare:2008ab, Adare:2009qk}. As this list shows, in contrast to ordinary atomic plasmas, the QGP is not observed directly, but rather through the hadronic (or the electromagnetic) residues, that are left after the transient QGP state expands, cools down and re-hadronizes~\cite{Stoecker:1986ci, Bass:1998vz}.

The distributions of hadrons produced in ultra-relativistic heavy-ion collisions indicate that the ``fireball'' created in the collision thermalizes, as they can be modelled very accurately using only a small number of parameters, such as temperature ($T$), chemical potential ($\mu$) and volume ($V$)~\cite{Becattini:2000jw, BraunMunzinger:2001ip}. The simplest theoretical model to describe this physics is a hadron-resonance gas~\cite{Hagedorn:1965st, Hagedorn:1980kb, Hagedorn:1984hz, Fiore:1984yu, Cleymans:1992zc}---see also ref.~\cite{Rafelski:2016hnq} for a historical account and ref.~\cite[section 2]{Wheaton:2004qb} for a modern overview of the main formul{\ae}. In its most elementary formulation (which continues to be a topic of active research to this day---see, e.g., ref.~\cite{Megias:2012hk} and references therein), it assumes that hadrons behave as an ideal gas of massive, free particles, and that their mutual interactions can be parameterized in terms of a tower of resonances~\cite{Dashen:1969ep}. In this picture, the pressure $p$ can be written as the sum of the contributions (denoted as $p_j$) from the different species of particles (labelled by $j$), which are assumed to be narrow, non-interacting, and to have finite mass $m_j$:
\begin{eqnarray}
\label{ideal_gas_pressure}
p &=& \sum_j p_j = \sum_j \frac{d_j}{6\pi^2} \int_0^{\infty} \dd k \frac{k^4 }{\sqrt{k^2+m_j^2}} \frac{1}{\exp \left[ \left. \left( \sqrt{k^2+m_j^2} - \mu_j \right) \right/ T \right]+\eta_j} \nonumber \\
&=& \frac{T^2}{2 \pi^2} \sum_j d_j m_j^2 \sum_{n=1}^\infty \frac{1}{n^2} \left[ \exp\left( \frac{n\mu_j}{T}\right) K_2\left( \frac{n m_j}{T} \right) - \frac{\eta_j+1}{4} \exp\left(\frac{2n\mu_j}{T}\right) K_2\left( \frac{2n m_j}{T}\right) \right],
\end{eqnarray}
where $d_j$ is the number of physical states (i.e. the spin degeneracy) for the generic particle species $j$, $\eta_j$ is $-1$ for bosons, while it equals $1$ for fermions, and  $K_\nu(z)$ denotes the modified Bessel function of the second kind of argument $z$ and index $\nu$. In general, each of the $p_j$ appearing in eq.~(\ref{ideal_gas_pressure}) is a function of the temperature $T$ and of the total chemical potential $\mu_j$ for the $j$th type of particles, which is defined as
\begin{equation}
\label{ideal_muj}
\mu_j = b_j \mu_B + q_j \mu_Q + s_j \mu_S,
\end{equation}
where $\mu_B$, $\mu_Q$ and $\mu_S$ denote the chemical potentials respectively associated with conservation of baryonic number, electric charge and strangeness, while $b_j$, $q_j$ and $s_j$ are the eigenvalues of these charges for the particle species $j$.

The success of this ``ideal'' hadron-resonance-gas model in describing the QCD thermodynamics in the confining phase is also confirmed by its comparison with numerical results from lattice calculations, which are based on the first principles of the theory, without any model-dependent assumptions. This has been shown for full QCD with dynamical fermions~\cite{Karsch:2003vd, Huovinen:2009yb, Borsanyi:2010cj, Borsanyi:2010bp, Bazavov:2011nk, Borsanyi:2013bia, Bazavov:2014pvz, Vovchenko:2014pka}, as well as for pure Yang-Mills theory\footnote{Note that, strictly speaking, the absence of quarks implies that no hadrons (neither baryons, nor mesons, nor other multi-quark states) exist in a purely gluonic theory. The spectrum of the theory only contains glueballs, i.e. color-singlet states made only of gluons. Throughout this article we nevertheless use a broad definition of ``hadrons'', which includes glueballs.}~\cite{Meyer:2009tq, Buisseret:2011fq, Borsanyi:2012ve, Caselle:2015tza}, and even in $2+1$ spacetime dimensions~\cite{Caselle:2011fy}. While the behavior of bulk thermodynamic quantities at very low temperatures is essentially accounted for by the contributions of the lightest hadron species only, heavier states start to play a more prominent r\^ole at temperatures $O(10^2)$~MeV: in particular, in Hagedorn's original picture~\cite{Hagedorn:1965st}, which predates QCD, an exponential growth in the hadron spectral density as a function of mass implies the existence of a limiting, maximal temperature at which hadronic matter can exist.

A possible way to improve the ideal hadron-resonance-gas model consists in including repulsive interactions through excluded-volume effects~\cite{Rischke:1991ke, Yen:1998pa, Yen:1997rv}: the idea is to assume that the total pressure $p$ is still given by the sum of the separate contributions from different particle species, like in eq.~(\ref{ideal_gas_pressure}), but that the $p_j$ are functions of $T$ and of a set of \emph{modified} chemical potentials $\mu_j^\star$ defined as
\begin{equation}
\label{modified_muj}
\mu_j^\star = b_j \mu_B + q_j \mu_Q + s_j \mu_S - v_j p,
\end{equation}
where $v_j$ denotes the ``eigenvolume parameter'' for the $j$th particle species. If the particles are modelled as hard spheres of radius $r_j$ and quantum-mechanical effects in their mutual hard-core interaction are neglected, then
\begin{equation}
\label{eigenvolume_parameter}
v_j = \frac{16 \pi}{3} r_j^3.
\end{equation}
Note, however, that quantum-mechanics effects are generally non-negligible~\cite{Kostyuk:2000nx}, so that $r_j$ should rather be interpreted as an ``effective radius''.

This excluded-volume-hadron-resonance-gas model has been used in recent comparisons with lattice results~\cite{Andronic:2012ut, Albright:2014gva, Vovchenko:2014pka}. Remarkably, in ref.~\cite{Vovchenko:2015cbk} it has been demonstrated that the chemical freeze-out temperature obtained in fits of experimental hadron yields is strongly sensitive to the hadron-volume parameters, i.e. to the details of the short-range repulsion between hadrons. The issue has been studied further in ref.~\cite{Alba:2016hwx}, in which different mass-volume relations were assumed for strange and non-strange hadrons: it was found that modelling the experimental results obtained in heavy-ion collisions by means of a gas of hadron resonances with excluded-volume effects yields much better fits of the observed particle distributions, if one assumes that heavier strange hadrons have smaller radii. In principle, also charmed and bottom mesons could show analogous behavior. Another recent study addressing related issues is ref.~\cite{Satarov:2016peb}, in which the equation of state is studied, under the assumption that mesons are point-like, while baryons and antibaryons have a finite hard-core radius. Refining the hadron-resonance-gas model for QCD could have important phenomenological implications: for example, it may improve the modelling of conserved-charge fluctuations~\cite{Garg:2013ata, Alba:2014eba, Nahrgang:2014fza, Luo:2014rea, Bhattacharyya:2015zka, Alba:2015iva, Bhattacharyya:2013oya, Alba:2015lxa}, which are an important tool to explore the QCD phase diagram.

In the present work, we extend the investigation of hadronic excluded-volume effects in a different direction---one that has the advantage of offering a somewhat ``clearer'' theoretical setup; namely we study this problem for the case of a purely gluonic theory, focusing, in particular, on the case of Yang-Mills theory with $\SU(2)$ gauge group, and carrying out a detailed comparison between the hadron-resonance-gas model with excluded-volume effects, and a novel set of continuum-extrapolated results from Monte~Carlo lattice simulations of this theory. In addition, we also present a similar analysis for the $\SU(3)$ Yang-Mills theory, whose equation of state has been determined in independent, high-precision calculations by various lattice groups~\cite{Boyd:1996bx, Umeda:2008bd, Giusti:2016iqr, Borsanyi:2012ve}; specifically, we use the data reported in ref.~\cite{Borsanyi:2012ve}.

The reasons why this type of study is interesting are manifold. From a purely theoretical side, pure Yang-Mills theory possesses only one (independent) physical scale---which can be chosen to be either the mass of the lightest state in the spectrum (a scalar glueball) or of some stable, heavier hadron, or the critical temperature $\Tc$ at which the second-order deconfinement transition takes place~\cite{Fingberg:1992ju, Engels:1994xj}, or the square root of the force between static fundamental color sources at asymptotically large distances, or the $\LambdaMSbar$ parameter of the theory, or any other dimensionful, non-perturbative scale: all of these quantities are related to each other by \emph{fixed} ratios, typically $O(1)$~\cite{Teper:1998kw, Lucini:2004my}. As a consequence, lattice simulations of pure Yang-Mills theory are \emph{completely} predictive, once one of these physical quantities has been chosen to set the scale, i.e. it has been assumed to take its experimentally measured value. This is a significant advantage with respect to QCD, where, in addition to the dimensionful scale generated non-perturbatively by quantum dynamics, all physical quantities exhibit dependence on parameters like the number of light quark flavors, their different masses, etc.: in particular, these parameters of the theory are known to affect significantly its finite-temperature properties~\cite{Karsch:2000kv}, in some cases even at the \emph{qualitative} level,\footnote{For example, the order and the very existence of a phase transition depends on the quark mass values~\cite{Brown:1990ev}.} and their effects on the physics can sometimes be difficult to disentangle from each other. In addition, Monte~Carlo simulations of purely bosonic theories have computational costs much lower than those of full lattice QCD with light dynamical quarks, and do not involve any of the subtleties related to the implementation of fermionic fields. As a consequence, one can obtain results of higher numerical precision; this is particularly important in the confining phase at low temperatures, in which the equilibrium thermodynamic observables take much smaller values than in the deconfined phase.

For a study of excluded-volume effects, focusing on a purely gluonic theory also entails an additional mathematical simplification: in a theory that contains no quarks, there is no baryonic number, no electric charge, and no strangeness either. As a consequence, all of the modified chemical potentials $\mu_j^\star$ defined in eq.~(\ref{modified_muj}) are simply proportional to the total pressure $p$.

Another, more ``phenomenological'', motivation to study the thermodynamic properties of pure Yang-Mills theory was put forward in ref.~\cite{Stoecker:2015zea}, in which it was pointed out that the early stages of the system produced in proton-proton, proton-nucleus and nucleus-nucleus collisions can be modelled by an essentially purely gluonic deconfined plasma. Some implications of this scenario and related aspects have been recently discussed in refs.~\cite{Beitel:2015bxa, Stocker:2015nka, Vovchenko:2015yia, Moreau:2015rrs, Beitel:2016ghw, Wunderlich:2016aed, Vovchenko:2016ijt, Feng:2016ddr, Vovchenko:2016mtf, Beitel:2016djh}.

We emphasize that the primary goal of this paper is not to propose a new way to describe Yang-Mills lattice data, but to test the existing excluded-volume description for hadronic interactions in QCD~\cite{Rischke:1991ke, Yen:1998pa, Yen:1997rv, Kostyuk:2000nx, Andronic:2012ut, Albright:2014gva, Vovchenko:2015cbk, Alba:2016hwx, Satarov:2016peb}, in theories with a different particle content. As we discuss in detail below, while purely gluonic $\SU(2)$ and $\SU(3)$ theories have many qualitative similarities with QCD, in some respects they are also remarkably different from it. As such, they can provide a useful testing ground to check the robustness of a model for hadron interactions in QCD at finite temperature, and give helpful indications as to what extent it can be reliably applied also for observables beyond equilibrium (like fluctuations of conserved charges) and/or in regions of the QCD phase diagram in which lattice calculations face challenges~\cite{deForcrand:2010ys, Gattringer:2016kco}.

The thermodynamics of Yang-Mills theory with $\SU(2)$ gauge group lends itself to testing excluded-volume corrections of the hadron-resonance-gas model in a setup with a non-trivial difference with respect to the $\SU(3)$ theory: the physical spectrum of the theory with $N=2$ colors does not contain any state of negative eigenvalue under the charge conjugation operator $\mathcal{C}$. This is a straightforward consequence of the fact that all irreducible representations of the algebra of the $\SU(2)$ group are real or pseudo-real, and implies a clear difference in the pressure of the theory at $T < \Tc$. See ref.~\cite[figure 4]{Caselle:2015tza}. Understanding how excluded-volume effects affect the thermodynamics of this theory, and comparing the results with the $\SU(3)$ Yang-Mills case may thus reveal interesting common patterns, and improve our understanding of such effects in full QCD, too. Another interesting feature of $\SU(2)$ Yang-Mills theory is that its deconfinement transition is of second order; hence the Hagedorn temperature $\THagedorn$ should be equal to the deconfinement temperature. As compared to the theory with $\SU(3)$ gauge group (in which the deconfinement transition is a weakly first-order one, and $\THagedorn > \Tc$), this removes a parameter from the fits, and strengthens the predictive power of the statistical-model description.

The structure of this article is the following. In section~\ref{sec:lattice} we set our notations and present the lattice formulation of the theory. In section~\ref{sec:results} we present the results of our Monte~Carlo simulations and their extrapolation to the continuum limit; these results are then analyzed and compared with a hadron-resonance-gas model (for which we use the glueball spectra previously determined in refs.~\cite{Teper:1998kw, Lucini:2004my}), studying excluded-volume effects. In section~\ref{sec:conclusions} we summarize our findings, discuss their implications for QCD, and list some future directions of research.

\section{Lattice setup}
\label{sec:lattice}

In this section, we introduce the definitions of the main quantities relevant for this work, and summarize the setup of our Monte~Carlo calculations, which is the same as in ref.~\cite{Caselle:2015tza}; we refer readers interested in technical details about the lattice calculation to that article, and to the earlier works mentioned therein.

We consider $\SU(2)$ Yang-Mills theory in a four-dimensional box of large, but finite, spatial volume $V=L_s^3$ and extent $L_t=1/T$ along the Euclidean-time direction, and regularize it on a lattice $\Lambda$ of spacing $a$, with $N_s=L_s/a$ sites along each spatial direction, and $N_t=L_t/a$ sites along the Euclidean-time direction. We define the Euclidean action of the lattice theory as~\cite{Wilson:1974sk}
\begin{equation}
\label{Wilson_action}
\Slat[U] = -\frac{2}{g^2} \sum_{x \in \Lambda} \sum_{0 \le \mu < \nu \le 3} \Tr U_{\mu\nu} (x),
\end{equation}
where $U_{\mu\nu} (x) = U_\mu (x) U_\nu \left(x+a\hat{\mu}\right) U_{\mu}^\dagger \left(x+a\hat{\nu}\right) U_{\nu}^\dagger (x)$ denotes the ``plaquette'', and $U_\mu (x)$ is the $\SU(2)$ matrix defined on the oriented link between nearest-neighboring sites $x$ and $x+a\hat{\mu}$. For later convenience, we define the Wilson action parameter $\beta=4/g^2$. We compute all expectation values of physical quantities by Monte~Carlo integration, using ensembles of configurations produced by an algorithm combining heat-bath~\cite{Creutz:1980zw, Kennedy:1985nu} and over-relaxation updates~\cite{Adler:1981sn, Brown:1987rra}, and estimate the statistical uncertainties of our simulation results by the jackknife method~\cite{bootstrap_jackknife_book}. We set the physical scale of our lattice simulations using the string tension $\sigma$ (in lattice units) extracted from the zero-temperature static quark-antiquark potential: for $2.25 \le \beta \le 2.6$, the values of $\sigma a^2$ for this theory can be interpolated by~\cite{Caselle:2016wsw}
\begin{equation}
\label{a_versus_beta}
\sigma a^2 = \exp \left[ - 2.68 - 6.82 \cdot (\beta -2.4) - 1.90 \cdot (\beta -2.4)^2 + 9.96 \cdot (\beta -2.4)^3 \right].
\end{equation}
Note that, for $\SU(2)$ Yang-Mills theory, the dimensionless ratio of the deconfinement critical temperature over the square root of the string tension is $\Tc/\sqrt{\sigma} = 0.7091(36)$~\cite{Lucini:2003zr}. 

Let us now define the main thermodynamic observables of the theory. In the canonical ensemble, the pressure $p$ is the intensive variable conjugate to the system volume, and, in the thermodynamic limit $V \to \infty$, equals minus the density of free energy $F$ per unit volume $f=F/V$:
\begin{equation}
\label{pressure}
p = -\lim_{V \to \infty} f = \lim_{V \to \infty} \frac{T}{V} \ln{Z}.
\end{equation}
The pressure is also related to the trace (denoted as $\Delta$) of the stress-energy tensor of the theory:
\begin{equation}
\label{trace_anomaly}
\frac{\Delta}{T^4} = T \frac{\partial}{\partial T} \left( \frac{p}{T^4} \right).
\end{equation}

Our lattice computation of these quantities is based on the integral method~\cite{Engels:1990vr}, using the $p=-f$ equality:\footnote{As stressed above, this equality is exact only in the infinite-volume limit; the finite-volume corrections to this equality have been studied in various articles~\cite{DeTar:1985kx, Elze:1988zs, Gliozzi:2007jh, Panero:2008mg, Meyer:2009kn} and turn out to have a negligible impact for the lattice sizes and temperatures investigated in the present work.} the calculation of the pressure is traded for the calculation of the free-energy density, which is proportional to $\ln Z$. In turn, this quantity is reconstructed by computing the derivative of $\ln Z$ with respect to $\beta$ (which is proportional to the expectation value of the trace of the average plaquette, $U_{\Box}$) and integrating it over $\beta$. The upper limit of this definite integral, to be denoted as $\beta^{(T)}$, is the value of the Wilson parameter yielding the lattice spacing that corresponds to the target temperature $T$, namely, $1 / \left[ N_t a\left( \beta^{(T)} \right) \right] = T$. The ultraviolet quantum fluctuations affecting this quantity are removed by subtracting from the integrand the expectation value of the trace of the plaquette calculated at the same $\beta$ (i.e. for the same lattice cutoff) on a lattice of sizes $\widetilde{N}^4$ (where $\widetilde{N}$ is sufficiently large, so that the temperature is approximately zero). Finally, we impose the condition that $\lim_{T\to0} p(T)=0$ by setting the lower integration limit to a value ($\beta^{(0)}$) at which the temperature of the system is close to zero, $1 / \left[ N_t a\left( \beta^{(0)} \right) \right] \simeq 0$. In summary, the pressure is obtained as
\begin{equation}
\label{integral_method}
p(T) = \frac{6}{a^4} \int_{\beta^{(0)}}^{\beta^{(T)}} \dd \beta \left[ \langle U_{\Box} \rangle_{\mathcal{T}(\beta)} - \langle U_{\Box} \rangle_0 \right],
\end{equation}
where the $\langle \dots \rangle_{\mathcal{T}(\beta)}$ notation denotes expectation value at the temperature $\mathcal{T}(\beta) = 1 / \left[ N_t a(\beta) \right]$. The right-hand side of eq.~(\ref{integral_method}) is computed by numerical integration of plaquette differences calculated at $n_\beta$ values of $\beta$ in the interval from $\beta^{(0)}$ to $\beta^{(T)}$, using the trapezoid method.\footnote{Other numerical integration methods~\cite[appendix]{Caselle:2007yc} give equivalent results, within the level of precision of our numerical data.}

The integrand in eq.~(\ref{integral_method}) is closely related to the trace of the stress-energy tensor,
\begin{equation}
\label{trace_anomaly_integral_method}
\Delta (T) = \frac{6}{a^4} \frac{\partial \beta}{\partial \ln a} \left( \langle U_{\Box} \rangle_0 -  \langle U_{\Box} \rangle_T \right),
\end{equation}
up to a factor that is obtained from the scale setting of the theory. Since $\Delta$ (unlike $p$) is evaluated directly on the lattice, in the following we focus on its behavior, comparing it with the hadron-resonance-gas model with excluded-volume effects. More precisely, we express the trace of the energy-momentum tensor in units of the fourth power of the temperature, and study it as a function of the ratio of the temperature $T$ over the deconfinement critical temperature $\Tc$.

\section{Lattice results and comparison with the hadron-resonance gas}
\label{sec:results}

\subsection{Results for the $\SU(2)$ theory}
\label{subsec:SU2_results}

Our results for the $\SU(2)$ theory are based on a set of lattice simulations at the parameters listed in table~\ref{tab:parameters}: this ensemble includes a part of the configurations analyzed in ref.~\cite{Caselle:2015tza}, and extends it with configurations on finer lattices, enabling us to extrapolate our results to the continuum limit.

\begin{table}
\begin{center}
\begin{tabular}{| c | c | c | c | c | c | c |}
\hline
$N_t$ & $N_s$ & $\widetilde{N}$ & $n_\beta$ & $\beta$ range & $\nconfig$ at finite $T$ & $\nconfig$ at $T=0$ \\
\hline \hline
 $6$ & $72$ & $40$ & $25$ & $[2.3059, 2.431]$ & $1.5 \times 10^5$ & $1.5 \times 10^5$ \\
 $7$ & $80$ & $40$ & $12$ & $[2.38,2.476]$ & $1.5 \times 10^5$ & $10^5$ \\
 $8$ & $80$ & $40$ & $14$ & $[2.42,2.516]$ & $1.5 \times 10^5$ & $10^5$ \\
$10$ & $96$ & $40$ & $12$ & $[2.51,2.58]$ & $6 \times 10^4$ & $10^5$ \\
\hline
\end{tabular}
\end{center}
\caption{Parameters of our lattice simulations of $\SU(2)$ Yang-Mills theory. The finite-temperature plaquette expectation values appearing on the right-hand side of eq.~(\protect\ref{trace_anomaly_integral_method}) are evaluated on lattices of sizes $N_t \times N_s^3$ (first two columns), while those at $T=0$ are obtained from simulations on lattices of sizes $\widetilde{N}^4$ (third column), at the same $\beta$ values. This is done at $\nbeta$ (fourth column) values of the Wilson parameter, in the interval reported in the fifth column. The statistics of thermalized, independent configurations in these runs is reported in the last two columns. This data sample includes part of the data used in ref.~\cite{Caselle:2015tza}.}
\label{tab:parameters}
\end{table}

\begin{figure}[h!]
\begin{center}
\includegraphics*[width=\textwidth]{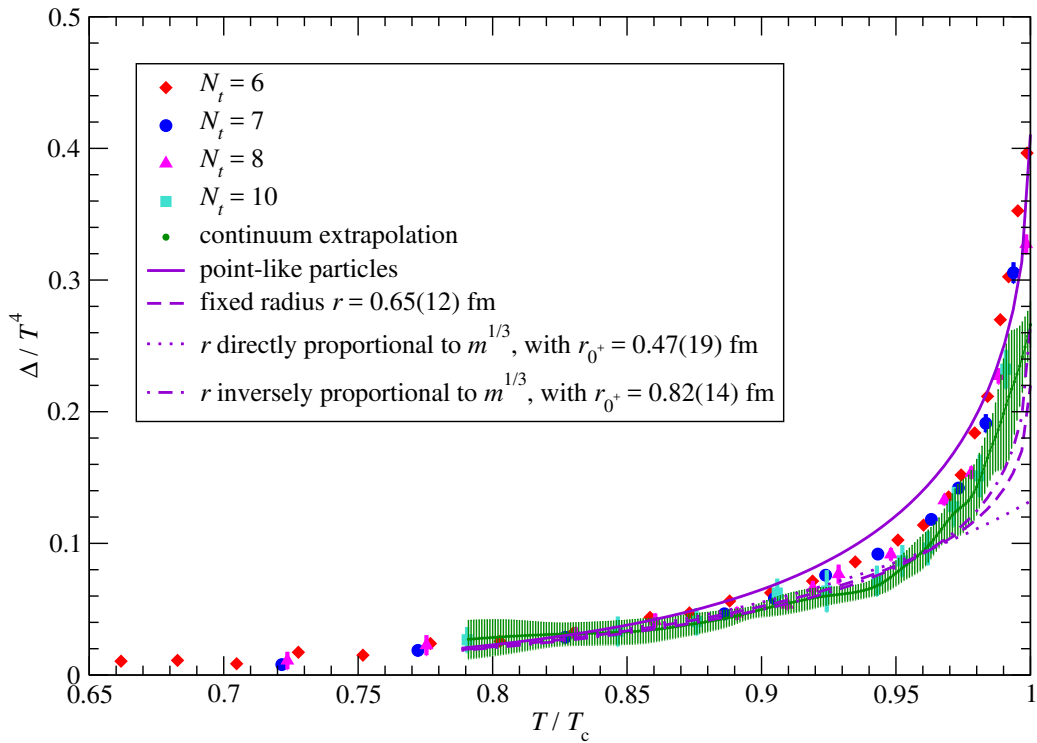}
\caption{\label{fig:SU2_Delta} Lattice results for the trace of the stress-energy tensor $\Delta$ (in units of $T^4$) in the confining phase of $\SU(2)$ Yang-Mills theory, from lattice simulations at $N_t=6$ (red diamonds), $7$ (blue circles), $8$ (magenta triangles) and $10$ (cyan squares), and their extrapolation to the continuum limit (green curve) with the associated error band. The results are plotted against the temperature $T$, in units of the critical deconfinement temperature $\Tc$. The figure also shows the fits of the hadron-gas model with or without excluded-volume effects (violet curves): the solid line is obtained under the assumption that particles are point-like, the dashed line assumes that all particles have the same radius, while the dotted line is based on the \emph{Ansatz} that the volumes of different glueballs are directly proportional to their mass, and finally the dash-dotted line is obtained assuming that the volume of each particle is inversely proportional to the particle mass. Information on these fits is summarized in table~\protect\ref{tab:SU2}.}
\end{center}
\end{figure}

The lattice results for $\Delta/T^4$, as a function of $T/\Tc$, are shown in fig.~\ref{fig:SU2_Delta}: symbols of different colors were obtained from simulations on lattices at different values of $N_t$ (from $6$ to $10$), i.e. at different lattice spacings. Fig.~\ref{fig:SU2_Delta} also displays the continuum extrapolation (green curve), which was constructed in the following way. First, at each $N_t$ value, we interpolated our results by cubic splines $f_{N_t}(T/\Tc)$. Then, we considered the values of these splines at each of the temperatures defined by
\begin{equation}
\label{interpolated_temperatures}
T_i = \left( 0.79 + i \times 10^{-3} \right) \Tc \qquad \mbox{for $i \in \N$, with $1 \le i \le 210$},
\end{equation}
and fitted each of them to
\begin{equation}
\label{continuum_extrapolation_form}
d_i(N_t) = d^{(0)}_i + \frac{d^{(1)}_i}{N_t^2}.
\end{equation}
The continuum-extrapolated value of $\Delta/T^4$ at the temperature $T_i$ was then defined as $d^{(0)}_i$. For each temperature $T_i$ defined in eq.~(\ref{interpolated_temperatures}), this procedure was then repeated on ten jackknife bins, in order to estimate the statistical error of the extrapolated result.

The main systematic uncertainties associated with this extrapolation procedure have two sources: the ambiguity in defining an interpolating form for data at fixed $N_t$, and the functional form to parameterize finite-cutoff corrections in eq.~(\ref{continuum_extrapolation_form}).

Concerning the interpolating form for the lattice data in each $N_t$ sample, our choice of a cubic spline is mainly motivated by the fact that this type of interpolation provides a general, minimal, smooth parameterization for the data, without specific assumptions about the functional form that should describe them. In addition, as compared, e.g., to polynomial interpolations or to Pad\'e approximants, it is well known that spline interpolation is not affected by the problem of Runge's phenomenon.

A rough estimate of the systematic uncertainty involved in the interpolation of our lattice data by a continuous curve can be obtained studying how much the results vary, if one uses a different interpolating function. To this purpose, we performed a polynomial interpolation of our data for each $N_t$, obtaining a curve that is in very good agreement with the result of the spline-interpolation procedure previously described. This suggests that the systematic uncertainty associated with the choice of an interpolating form is indeed under control, and much smaller than the statistical uncertainty of our results.

Similarly, the ambiguity associated with the continuum extrapolation can be estimated, by carrying out such extrapolation using a functional form different from eq.~(\ref{continuum_extrapolation_form}). Given that the finite-lattice-spacing artifacts affecting the action and the observables in our lattice formulation are expected to be proportional to powers of $a^2$ (hence to powers of $1/N_t^2$, when the temperature $T$ is fixed), one could estimate the systematic error associated with the continuum extrapolation by including a further addend $d^{(2)}_i/N_t^4$ on the right-hand side of eq.~(\ref{continuum_extrapolation_form}), fitting also $d^{(2)}_i$ (in addition to $d^{(0)}_i$ and $d^{(1)}_i$), and defining the continuum-extrapolated value of $\Delta/T^4$ at that temperature as the $d^{(0)}_i$ coefficient obtained from this three-parameter fit. However, this procedure eventually leads to a much less stable continuum extrapolation, because the $d^{(2)}_i$ coefficient turns out to be poorly determined. In particular, the resulting curve is very sensitive to the $N_t=10$ data, and their comparatively large uncertainties ultimately lead to unphysical results: hence, for the continuum extrapolation shown in fig.~(\ref{fig:SU2_Delta}) we chose the functional form of eq.~(\ref{continuum_extrapolation_form}), without considering any further powers of $1/N_t^2$.

Since interactions among glueball states are poorly known~\cite{Mathieu:2008me}, we perform our analysis of the lattice data testing the different parameterizations for the particle eigenvolume that were already discussed in ref.~\cite{Alba:2016hwx}, with \emph{fixed} radius
\begin{equation}
\label{fixed}
r_j = r_{0^+}, \qquad \forall j
\end{equation}
(where $j$ labels the glueball state, and $r_{0^+}$ denotes the radius of the lightest glueball, with quantum numbers $J^{\mathcal{P}}=0^+$ and mass $m_{0^+}$), with volume \emph{directly} proportional to the glueball mass $m_j$, which implies 
\begin{equation}
\label{direct}
r_j = \sqrt[3]{\frac{m_j}{m_{0^+}}} r_{0^+},
\end{equation}
or with volume \emph{inversely} proportional to the mass of the particle, i.e.
\begin{equation}
\label{inverse}
r_j = \sqrt[3]{\frac{m_{0^+}}{m_j}} r_{0^+}.
\end{equation}
Thanks to the high quality of the data, we are able to test the physical assumptions of these parameterizations, which, for real-world QCD, turn out to have a strong impact on the description of experimental data~\cite{Alba:2016hwx, Vovchenko:2016ebv}. Besides the simplest scenario described by eq.~(\ref{fixed}), in which all particles have the same radius, eq.~(\ref{direct}) and eq.~(\ref{inverse}) respectively describe the possibility that the eigenvolume increases or decreases with the particle mass. Even though this could seem unjustified, it is presently not clear how higher-mass resonances in a certain channel would interact, as compared with the ground state; in general, there is the possibility that they may have a smaller cross section, which would be encoded in a smaller effective radius. This could be particularly relevant for a correct inclusion of exotic resonances, for which the repulsive channels are known to be as relevant as the attractive ones~\cite{Friman:2015zua}. In the case of mass-dependent eigenvolumes, we label the parameterization in terms of the radius of the ground-state $J^{\mathcal{P}}=0^+$ particle, in order to have an immediate comparison to the fixed-radius scenario.

As shown in various recent works~\cite{Meyer:2009tq, Buisseret:2011fq, Borsanyi:2012ve, Caselle:2015tza}, the contribution from a tower of Hagedorn states to pure Yang-Mills thermodynamics is non-negligible. In the present analysis, we consider the same Hagedorn spectrum used in ref.~\cite{Caselle:2015tza}, which is expected to model the spectrum states with mass larger than twice the mass of the lightest glueball (i.e. larger than $3291.2$~MeV), and $\THagedorn=\Tc=0.7091(36)\sqrt{\sigma} \simeq 312(2)$~MeV. The fits are performed minimizing the $\chi^2$ per degree of freedom (that we denote as $\redchisq$). Although our extrapolated curves yield the continuum value of $\Delta/T^4$ at any $T$ in the temperature interval from $0.79 \Tc$ to $\Tc$, it is clear that, by construction, the $\Delta/T^4$ at nearby temperature values (and their uncertainties) are strongly correlated with each other. In order to define a fitting procedure that bypasses the complications generated by such spurious correlations, we computed $\redchisq$ using the continuum-extrapolated lattice data evaluated at only twenty, equally spaced, values of $T/\Tc$ within the temperature interval in which the continuum-extrapolated curve is defined; this set of temperatures is approximately the same as the actual set of temperature values probed in independent lattice simulations. We stress that all of the fits were performed at temperatures strictly less than $\Tc$. The uncertainty on the fitted parameters was obtained imposing the $\chi^2+1$ criterion (see, e.g., ref.~\cite{Andronic:2016nof}).

In table~\ref{tab:SU2} we summarize our fit results, which are shown by the violet curves in fig.~\ref{fig:SU2_Delta}. As compared to the model with point-like particles, it is clear that the inclusion of short-range repulsions dramatically improves the quality of the hadron-resonance-gas description, with reasonable values of the glueball radii. In fig.~(\ref{fig:SU2_Delta}) we compare the curves obtained from the different fits (and the curve based on the \emph{Ansatz} of point-like glueballs), assuming that the particles have the same, finite, eigenvolume, or that they have an eigenvolume directly or inversely proportional to their mass: it is interesting to observe that the assumption of a glueball volume proportional to the inverse of the mass yields the best data description among the three ways of modelling excluded-volume effects, that we considered.

\begin{table}[h]
\begin{center}
\begin{tabular}{|l|c|c|c|c|}
\hline
volume-mass dependence  & $r_{0^+}$ (fm) & $\delta r_{0^+}$ (fm) & $\redchisq$ \\
\hline
\hline
point-like particles    & $0$           & $0$            & $8.16$     \\
constant radius         & $0.65$        & $0.12$         & $0.74$     \\
direct proportionality  & $0.47$        & $0.19$         & $1.87$      \\
inverse proportionality & $0.82$        & $0.14$         & $0.39$      \\
\hline
\end{tabular}
\end{center}
\caption{Best-fit results of our lattice data for the $\SU(2)$ interaction measure, to the glueball-gas model. The radius, with the corresponding error, of the lightest glueball state (the ground-state particle in the channel with quantum numbers $J^{\mathcal{P}}=0^+$) and the $\redchisq$ value are shown for different scenarios.}
\label{tab:SU2}
\end{table}

\begin{figure}[h!]
\begin{center}
\includegraphics*[width=\textwidth]{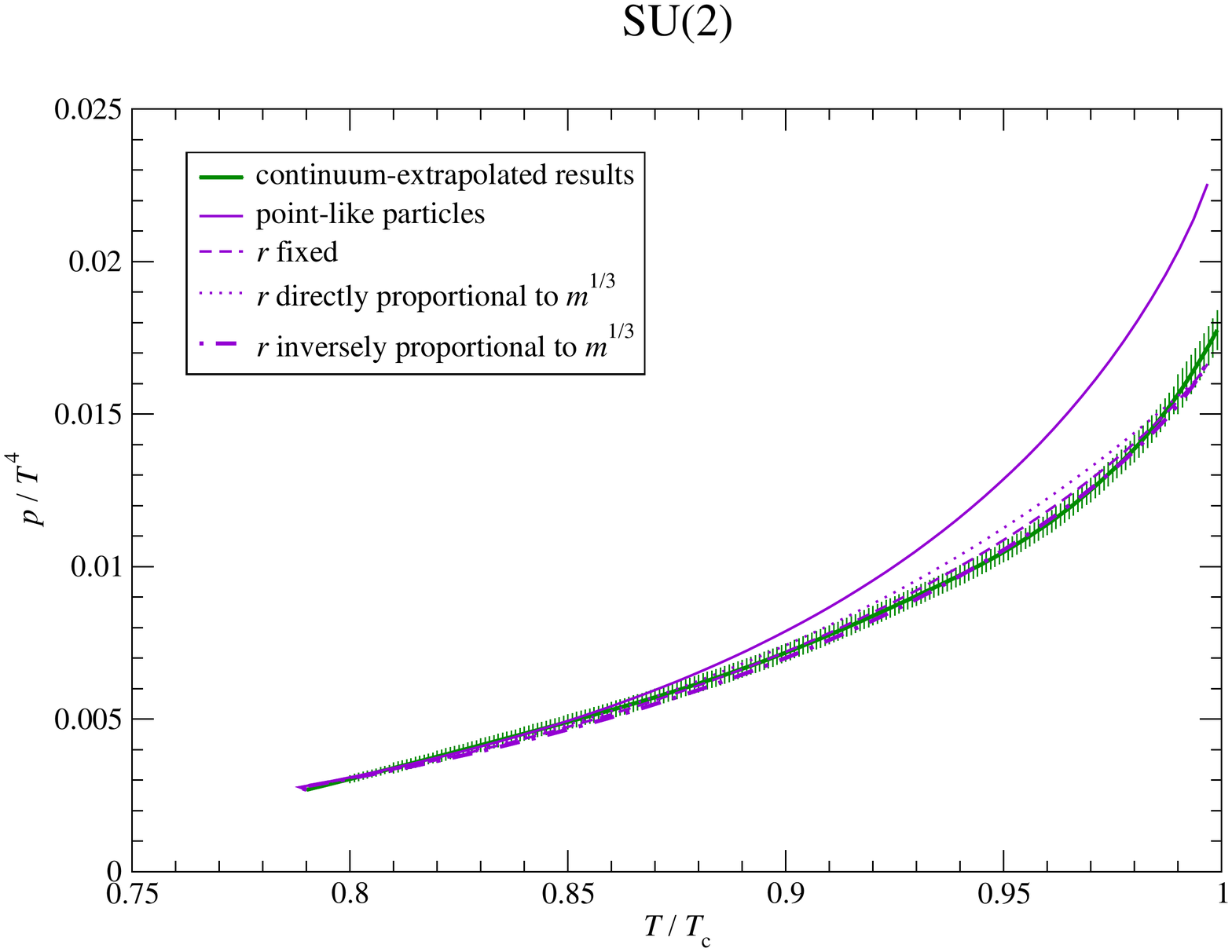}
\caption{\label{fig:SU2_pressure} Pressure (in units of $T^4$) in $\SU(2)$ Yang-Mills theory: the figure shows a comparison of our continuum-extrapolated lattice results (green line) and the hadron-resonance-gas predictions obtained by integration of the different fits of $\Delta/T^4$ in fig.~\protect\ref{fig:SU2_Delta}, assuming point-like particles (violet solid line), particles of constant volume (violet dashed line), particles with eigenvolume directly proportional to their mass (violet dotted line), or particles with eigenvolume inversely proportional to their mass (violet dash-dotted line). All curves are obtained using eq.~(\protect\ref{trace_anomaly}), with the integration constant $p/T^4=0.00268$ for $T/\Tc=0.79$~\cite{Caselle:2015tza}.}
\end{center}
\end{figure}

Our results for the pressure in $\SU(2)$ Yang-Mills theory are shown in fig.~\ref{fig:SU2_pressure}.

Before discussing the $\SU(3)$ theory, we mention one additional observation. It is interesting to investigate what happens, if one compares the lattice data for $\Delta/T^4$ with two-parameter fits, in which also $\mthreshold$, besides $r_{0^+}$, is fitted. In general, this determines these quantities very poorly: for example, assuming the particles to have the same radius, one finds $r_{0^+}=0.5(4)$~fm and $\mthreshold=4100(1600)$~MeV, with $\redchisq \simeq 0.4$. If the dependence between the particles' radii and their masses is of the form in eq.~(\ref{direct}), then one finds $r_{0^+}=0.2(3)$~fm and $\mthreshold=4700(1100)$~MeV, again with $\redchisq \simeq 0.4$. Finally, if the volume is taken to be inversely proportional to the particle mass, one obtains $r_{0^+}=0.7(5)$~fm and $\mthreshold=3600(1900)$~MeV, and again $\redchisq \simeq 0.4$. The large uncertainties on both $r_{0^+}$ and $\mthreshold$, as well as the rather small $\redchisq$ values, indicate that this type of analysis tends to ``overfit'' the lattice data.

\subsection{Results for the $\SU(3)$ theory}
\label{subsec:SU3_results}

In order to further check our assumption, we performed the same analysis for the $\SU(3)$ data from ref.~\cite{Borsanyi:2012ve}. In this case, for the Hagedorn temperature we assumed the value $\THagedorn=1.024 \Tc$ found in ref.~\cite{Meyer:2009tq} and used also in ref.~\cite{Borsanyi:2012ve}. In table~\ref{tab:SU3} we show the results of our analysis. Similarly to the $\SU(2)$ case, we found that including eigenvolume effects yields a significant improvement of the hadron-resonance-gas description of data: this is mainly due to the points close to the transition. The glueball radii are comparable to those obtained for the two-color theory, and the quality of the description with excluded-volume effects is roughly the same for the three different types of mass-volume dependence, but the fit assuming glueball volumes directly proportional to the masses is clearly worse than the other two---while the one assuming that the eigenvolume of each particle is inversely proportional to its mass is the best one, with $\redchisq \simeq 1$.

\begin{figure}[h!]
\begin{center}
\includegraphics*[width=\textwidth]{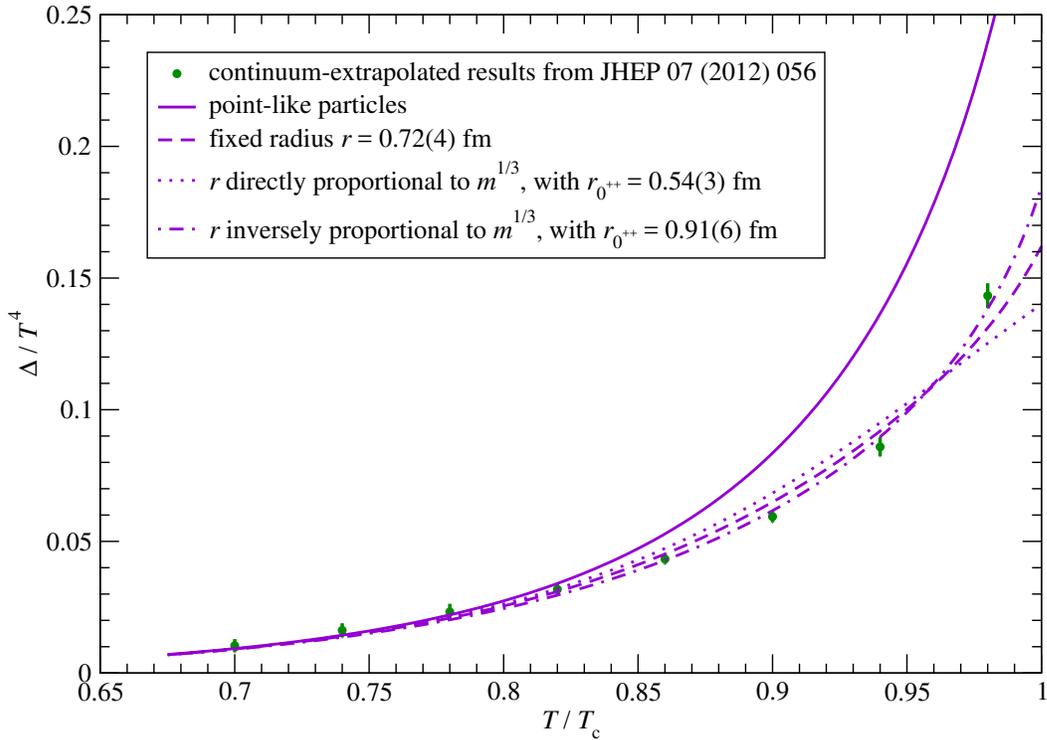}
\caption{\label{fig:SU3_Delta} Same as in fig.~\protect\ref{fig:SU2_Delta}, but for the continuum-extrapolated lattice results from ref.~\cite{Borsanyi:2012ve} for $\SU(3)$ Yang-Mills theory, assuming $\THagedorn = 1.024 \Tc$~\cite{Meyer:2009tq, Borsanyi:2012ve}. These fits are summarized in table~\protect\ref{tab:SU3}.}
\end{center}
\end{figure}

\begin{table}[h]
\begin{center}
\begin{tabular}{|l|c|c|c|c|}
\hline
volume-mass dependence  & $r_{0^{++}}$ (fm) & $\delta r_{0^{++}}$ (fm) & $\redchisq$ \\
\hline
\hline
point-like particles    & $0$               & $0$             & $84.3$     \\
constant radius         & $0.733$           & $0.08$          & $2.33$      \\
direct proportionality  & $0.55$            & $0.07$          & $5.41$      \\
inverse proportionality & $0.91$            & $0.10$          & $0.82$      \\
\hline
\end{tabular}
\end{center}
\caption{Results of the best fit on the lattice data for the $\SU(3)$ interaction measure from ref.~\cite{Borsanyi:2012ve}. The radius of the lightest glueball state (with quantum numbers $J^{\mathcal{PC}}=0^{++}$) and its uncertainty are shown for different scenarios, together with the corresponding $\redchisq$ values.}
\label{tab:SU3}
\end{table}

\begin{figure}[h!]
\begin{center}
\includegraphics*[width=\textwidth]{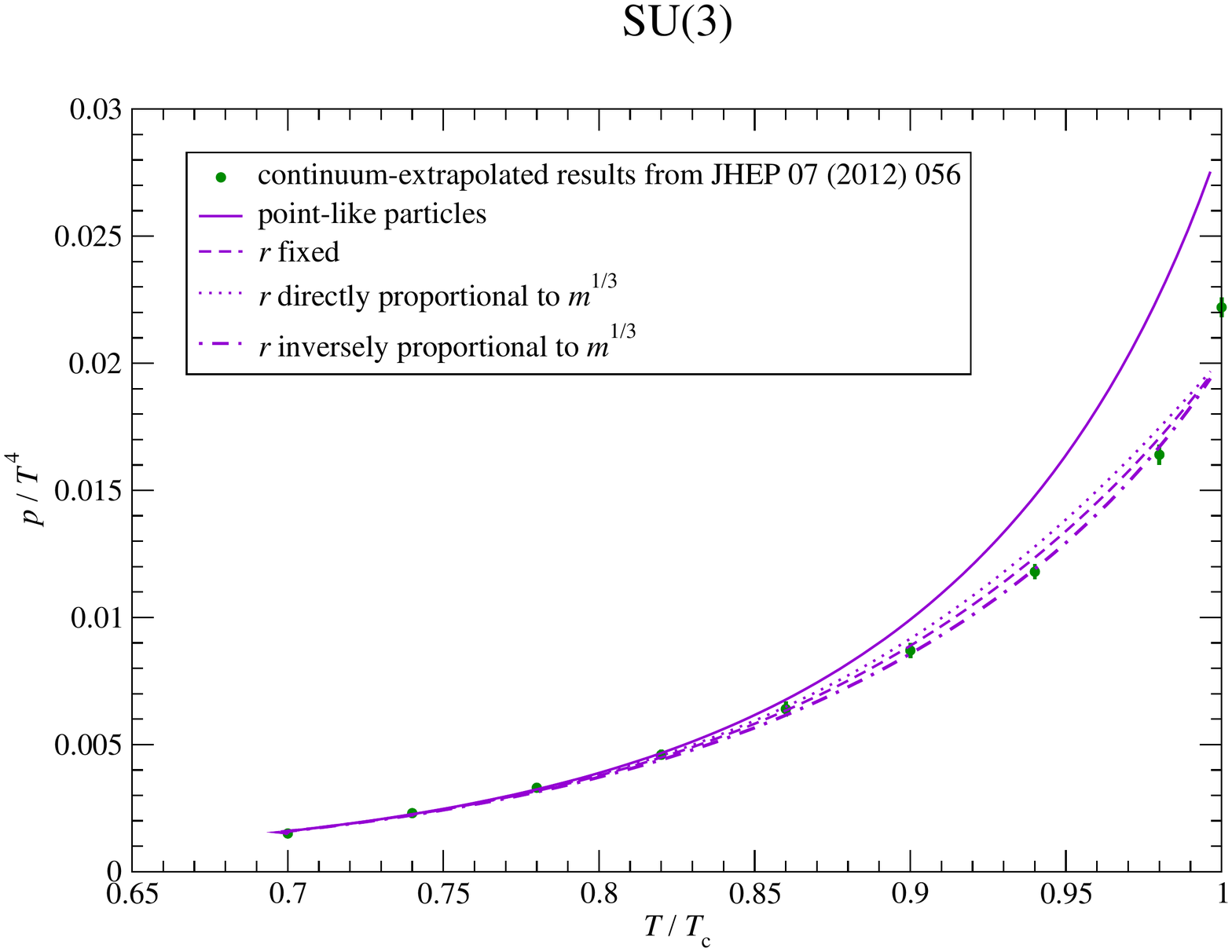}
\caption{\label{fig:SU3_pressure} Same as in fig.~\protect\ref{fig:SU2_pressure}, but for $\SU(3)$ Yang-Mills theory: the plot shows a comparison of the lattice results from ref.~\cite{Borsanyi:2012ve} (green symbols) and the statistical-model predictions (violet lines) obtained by integration of the curves shown in fig.~\protect\ref{fig:SU3_Delta}, assuming the integration constant $p/T^4=0.0015(1)$ for $T/\Tc=0.7$~\cite{Borsanyi:2012ve}.}
\end{center}
\end{figure}

Figure~\ref{fig:SU3_pressure} shows a comparison of the results for the pressure in $\SU(3)$ Yang-Mills theory obtained in ref.~\cite{Borsanyi:2012ve} with the curve obtained by integration of our result for the hadron-resonance-gas model with and without excluded-volume effects, for the different types of relations between the particle eigenvolume and mass.

Finally, carrying out two-parameter fits, in which both $r_{0^{++}}$ and $\mthreshold$ are regarded as free parameters, one obtains $r_{0^{++}}=0.68(6)$~fm, $\mthreshold=3200(300)$~MeV, and $\redchisq \simeq 1.3$ when glueballs are assumed to have a common radius, $r_{0^{++}}=0.46(7)$~fm, $\mthreshold=3400(300)$~MeV, and $\redchisq \simeq 1.8$ when the glueball volume is assumed to be directly proportional to their mass, and $r_{0^{++}}=0.9(1)$~fm, $\mthreshold=3000(300)$~MeV, and $\redchisq \simeq 0.8$ when the glueball volume is assumed to be inversely proportional to their mass.

\section{Discussion and conclusions}
\label{sec:conclusions}

In this work, we extended recent studies of excluded-volume effects in the hadron-resonance-gas model to the purely gluonic $\SU(2)$ gauge theory, whose continuum equation of state in the confining phase was determined by means of a novel set of high-precision lattice simulations. Our continuum-extrapolated results for this theory reveal some non-negligible deviations with respect to those reported in ref.~\cite{Caselle:2015tza} (in which no continuum extrapolation was attempted) in the temperature region closest to $\Tc$. This is not surprising, since the finite-lattice-spacing data reported in ref.~\cite{Caselle:2015tza} already revealed that the $N_t=8$ data for $\Delta/T^4$ in that temperature range are clearly lower than those obtained from $N_t=6$. Part of the motivation of the present work consisted in carrying out a reliable continuum extrapolation for the $\SU(2)$ data in the confining phase.

Then, we also carried out the analysis of excluded-volume effects for the $\SU(3)$ theory.

As is well known, the thermodynamic properties of Yang-Mills theories based on different gauge groups are expected to have a different dependence on the number $N$ of color charges in the confining and in the deconfined phases. At very high temperatures, color liberation and asymptotic freedom imply that the pressure is proportional to the number of physical gluon degrees of freedom, i.e. to $\npol \cdot \dadj$, where $\npol$ is the number of transverse polarizations ($2$ in $3+1$ spacetime dimensions) and $\dadj$ is the dimension of the adjoint representation of the gauge group algebra ($N^2-1$ for $\SU(N)$ gauge group): this is indeed confirmed by lattice calculations~\cite{Bringoltz:2005rr, Panero:2009tv, Datta:2010sq, Bruno:2014rxa}, even at temperatures very close to $\Tc$ (where the plasma is very different from a gas of free gluons), and even in $2+1$ spacetime dimensions~\cite{Caselle:2011mn}. By contrast, the physical degrees of freedom in the confining phase are hadrons, i.e. color-singlet states, whose number is $O(N^0)$. Nevertheless, the number of hadronic states (glueballs) in $\SU(2)$ Yang-Mills theory is different from the other $\SU(N \ge 3)$ theories, because purely group-theoretical facts imply that in the $\SU(2)$ theory no $\mathcal{C}=-1$ states can be formed. This reduced number of physical degrees of freedom implies that the equation of state in the confining phase of the $\SU(2)$ theory is significantly different with respect to the $\SU(3)$ theory. Moreover, the fact that the $\SU(2)$ Yang-Mills theory is quite ``simple'' from a conceptual (the physics depends only on one dimensionful scale) and computational (a purely bosonic, local theory, whose elementary degrees of freedom in the lattice regularization can be represented in a very compact form by pairs of complex numbers) point of view, makes it an ideal benchmark to study the hadron-resonance-gas model and the effect of excluded-volume corrections.

The analysis that we carried out in this work shows the improvement in the description of pure-glue thermodynamics, when repulsive interactions among glueball states are accounted for. As tables~\ref{tab:SU2} and~\ref{tab:SU3} show, this improvement is very clear for both the $\SU(2)$ and $\SU(3)$ theories. Our analysis also shows that two-parameter fits, in which both $r_{0^{++}}$ and $\mthreshold$ are considered as free parameters, tend to overfit the lattice data; nevertheless, at least for the $\SU(3)$ theory, they still tend to favor the assumption of glueball radii inversely proportional to their masses.

In short, we found that the effective sizes of the glueballs are \emph{finite}, consistent among the two theories, and slightly larger than the ones usually found in QCD. This could imply that the repulsive channels for glueball interactions are stronger than those for mesons and baryons. We also found that, both for $\SU(2)$ and $\SU(3)$ Yang-Mills theory, the best fits to the trace of the energy-momentum tensor (and to the pressure, which is directly linked to it) are obtained if one assumes that the eigenvolume of different hadronic states is inversely proportional to their mass. This is consistent with the results obtained for QCD in ref.~\cite{Alba:2016hwx}.

Some aspects of our present analysis (and their relation to previous works) deserve comments. First of all, one should note that, in QCD, eq.~(\ref{modified_muj}) is just a modification of the total chemical potential for the particle species $j$, eq.~(\ref{ideal_muj}), while in the purely gluonic theory glueballs carry no baryon number, no electric charge, no strangeness, so that the only non-vanishing contribution to the modified chemical potential $\mu_j^\star$ is the one arising from the excluded-volume term. As a consequence, one could wonder whether eq.~(\ref{modified_muj}) is the only (or the most appropriate) way to introduce finite-eigenvolume corrections parameterizing the effects of particle interactions in Yang-Mills theory. \emph{A priori}, there is no reason to assume that eq.~(\ref{modified_muj}) provides the only way to study excluded-volume effects in a glueball gas, but this approach has the notable advantage of allowing a direct comparison with the existing results obtained in QCD using the same method. As we discussed above, our results show a consistent pattern in the $\SU(2)$ and $\SU(3)$ Yang-Mills theories, and in QCD: finite-eigenvolume effects lead to a significant improvement of fits to the hadron-resonance-gas model, the particle radii fitted in the three theories are comparable, and are consistent with the same type of dependence on the particle mass.

For the $\SU(2)$ theory, it is worth remarking that, close to $\Tc$, the continuum-extrapolated curve is systematically (and significantly) lower than the $N_t=6$ and $N_t=8$ lattice data, and, in contrast to the latter, exhibits quite a clear deviation from the curve based on the Hagedorn model with no finite-volume effects, which overshoots it for all $T \gtrsim 0.87 \Tc$. The continuum extrapolation of $\SU(2)$ data carried out in this work enabled us to reveal this feature, and to identify an interpretation for it, in terms of an excluded-volume effect. Note that, for this theory, there is no ambiguity in defining the value of $\THagedorn$.

Although the idea that heavier hadrons have smaller radii has already been suggested in the theoretical literature~\cite{Friedmann:2009mx, Friedmann:2009mz, Friedmann:2012kr} and is supported by experimental evidence~\cite{Belle:2011aa, Olive:2016xmw}, our finding of eigenvolume values \emph{inversely} proportional to the particle mass may appear at odds with intuition, and in contrast to the expectations from simple semi-classical models~\cite{Isgur:1984bm}. In addition, it is perhaps worth mentioning that some lattice studies~\cite{Loan:2006gm, Liang:2014jta} found indication that heavier glueball states tend to have better overlap with more extended (rather than more localized) operators. We do not think that these results are necessarily in contradiction with our findings, because, as we already remarked, the eigenvalue parameters that we fitted are not to be interpreted as strictly equivalent to the physical volume of each state: rather, they account for the effects of glueball interactions, and describe the effective volume of each state. This means that some of the parameters that we fitted could turn out to be small, just because the corresponding types of glueball are weakly interacting, regardless of their actual physical size. Moreover, even if one assumed the fitted effective eigenvolumes to coincide with the physical volumes of the glueballs, it should be noted that our fits do not allow one to determine the precise volume of each particle to very high precision. Indeed, the three finite-eigenvolume scenarios that we considered here (i.e. the one in which the glueballs are assumed to have a common finite volume, the one in which their volume is directly proportional to their mass, and the one in which it is inversely proportional to the particle mass), described by eqs.~(\ref{fixed}), (\ref{direct}) and~(\ref{inverse}), are, at best, crude idealizations; in particular, they completely neglect the non-trivial non-perturbative dynamics accounting for the very existence of these states. Nevertheless, in our analysis we assumed these simple scenarios, in order to limit the number of parameters to be fitted to a minimum, and to try and capture at least the main features of the relation between the particle eigenvalue and mass. While our data indicate that an inverse proportionality relation between eigenvolume and mass provides the best fit to the data, the results in tables~\ref{tab:SU2} and~\ref{tab:SU3} (and the curves shown in figs.~\ref{fig:SU2_Delta} and~\ref{fig:SU3_Delta}) reveal that the other two finite-eigenvolume scenarios are not dramatically worse. The main reason for this ambiguity lies in the fact, that the dominant contribution to the thermodynamics comes from the lightest state in the spectrum, while those from heavier states are exponentially suppressed. This makes it particularly hard to distinguish, whether interactions involving heavy states are best described in terms of a fixed, increasing, or decreasing effective volume. In any case, the fits that we performed in this work provide clear evidence that the model with point-like (i.e. non-interacting) particles is ruled out, and that the radius of the lightest glueball is in the ballpark of $0.5$--$0.9$~fm (depending on the fit details). In particular, for the $\SU(2)$ theory, the present data analysis, performed on results extrapolated to the continuum limit, supersedes the more qualitative study presented in ref.~\cite{Caselle:2015tza}, in which the precision and accuracy of data sets at finite lattice cutoff was sufficient to confirm that heavy states give a large contribution to the equation of state for $T \gtrsim 0.8 \Tc$, but did not allow a continuum extrapolation or a $\chi^2$ analysis of effects beyond the simplest, point-like-particle, picture.

As another technical comment, it is also worth mentioning that, over the years, several lattice computations of glueball spectra have been reported~\cite{Berg:1982kp, Bali:1993fb, Morningstar:1997ff, Teper:1998kw, Morningstar:1999rf, Lucini:2004my, Meyer:2004gx, Chen:2005mg, Lucini:2010nv}: they are based on numerical calculations that differ in some technical aspects, and their results exhibit some quantitative discrepancies with each other. Repeating our analysis using spectra from different lattice studies (restricting our attention to the recent studies presented in refs.~\cite{Teper:1998kw, Lucini:2004my, Meyer:2004gx, Chen:2005mg}) leads to modest quantitative differences in the results, without changing them at a qualitative level.

Another interesting question concerns the robustness of the results obtained with a stringy Hagedorn spectrum against a different choice for the value of the lowest end $\mthreshold$ of the continuous part of the spectrum---see ref.~\cite[eq.~(3.8)]{Caselle:2015tza}. While setting $\mthreshold$ to twice the mass of the lightest particle in the physical spectrum may be regarded as the most natural choice (and in this work we stick to that choice), the assumption that the spectrum can be exactly split into a discrete set of light states, plus a continuum that is described by a bosonic-string model, is a crude approximation at best, and the very existence of a sharp threshold value $\mthreshold$ separating the two parts of the spectrum is an idealization. As a consequence, one may wonder, how the results would vary, should one choose different values of $\mthreshold$. We observe that our results are quite robust under a change in $\mthreshold$. In particular, they are essentially stable if $\mthreshold$ is varied to $3$, $3.3$, or $4$~GeV: this is consistent with the fact that the lightest states are those that contribute most to the thermodynamics. Somewhat larger variations are observed when $\mthreshold$ is reduced down to values that are significantly lower than twice the lightest glueball mass, but the results are affected in a strong way only for $\mthreshold \simeq 1$~GeV; obviously, however, the latter value is grossly unphysical, since it is even lower than the mass of the lightest glueball.

Some readers may wonder if including the contributions from the continuous part of the spectrum is just a way of modelling a poor knowledge of the number and masses of glueballs below $\mthreshold$. This is not so: the identification of states lighter than $\mthreshold$ from recent lattice calculations is unambiguous, and the level of precision to which their masses are known~\cite{Teper:1998kw, Lucini:2004my, Meyer:2004gx, Chen:2005mg} is sufficient to rule out the hypothesis that they may account for the thermodynamics. With the exception of the lightest states, the contribution of such glueballs to the pressure of the system is basically \emph{negligible}: the lattice data confirm that, exactly as in Hagedorn's original intuition~\cite{Hagedorn:1965st}, as $T \to \Tc^-$, the thermodynamics can only be reproduced in terms of contributions from a continuous, exponentially increasing density of states---whereby the growing spectral multiplicity (over-)compensates the exponential suppression of heavier and heavier particles.

One may also wonder up to which temperature a hadron gas is expected to model the Yang-Mills thermodynamics accurately. For the $\SU(2)$ theory, this question has a clear answer: the description in terms of a gas of massive hadrons must fail at a temperature strictly lower than $\Tc$, because the deconfinement transition is a continuous one, and the dynamics in the proximity of $\Tc$ should be characterized by the critical exponents of the three-dimensional Ising model~\cite{Svetitsky:1982gs}---an expectation that is indeed borne out by lattice calculations~\cite{Engels:1989fz, Velytsky:2007gj}. Moreover, as we already pointed out in sec.~\ref{sec:intro}, the fact that the deconfining transition in $\SU(2)$ Yang-Mills theory is of second order also implies that $\Tc$ equals the Hagedorn temperature $\THagedorn$, at which, according to the statistical model, the spectral density (and the bulk thermodynamic observables) would diverge, but no such divergence is observed in lattice simulations. Clearly, this signals the breakdown of the hadron-resonance-gas model---rather than the existence of an actual ``ultimate temperature''---and the transition to another state of matter, characterized by quantitatively different degrees of freedom, i.e. deconfined gluons. As a consequence, the temperature range in which a hadron-gas description holds must necessarily be limited to a finite temperature strictly less than $\Tc$. Nevertheless, one may wonder whether \emph{some} hadrons survive in the deconfined phase. The question is non-trivial, and entails deep phenomenological implications. We remark, however, that a full-fledged lattice study of the survival of glueballs in the deconfined phase would require the investigation of the temperature dependence of the appropriate spectral functions, which would likely be even more challenging than those for quarkonia~\cite{Asakawa:2003re, Datta:2003ww, Aarts:2007pk, Aarts:2011sm}, and which is clearly beyond the scope of this work. In our present analysis, we note that our results remain consistent within the uncertainties, when we restrict the temperature range to $T < 0.99 \Tc$, $T < 0.95 \Tc$, or even $T < 0.9 \Tc$ (in the latter case, however, the analysis tends to lose sensitiveness to excluded-volume effects).

Finally, it is worth discussing the implications of a different value for the Hagedorn temperature $\THagedorn$. In particular, in ref.~\cite{Caselle:2015tza} it was pointed out that the effective string model provides an excellent quantitative description of $\SU(3)$ Yang-Mills thermodynamics, if one uses the prediction for $\THagedorn$ from the Nambu-Got\={o} model (that is, the temperature at which the effective string tension predicted by the Nambu-Got\={o} string model vanishes),
\begin{equation}
\label{Nambu_Goto_THagedorn}
\THagedorn=\sqrt{\frac{3 \sigma}{2\pi}}.
\end{equation}
Numerically, this value corresponds to $\THagedorn \simeq 1.098\Tc$, which is significantly larger than the value used in refs.~\cite{Meyer:2009tq, Borsanyi:2012ve}, $\THagedorn = 1.024(3)\Tc$. As discussed in ref.~\cite{Meyer:2009tq}, the latter value was obtained by determining the temperature at which the inverse correlation length of the temporal flux loop, extracted from a two-point Polyakov-loop correlation function in $\SU(3)$ Yang-Mills theory, vanishes. Using this value for the Hagedorn temperature, in ref.~\cite{Meyer:2009tq} it was shown that the hadron-resonance-gas model (with point-like particles) yields a good description of the equation of state, provided the ``continuous'' part of the spectrum is modelled in terms of a closed bosonic string (neglecting the possibility of degenerate $\mathcal{C}=-1$ states). This analysis was later extended in ref.~\cite{Caselle:2015tza}, where it was pointed out that, if one includes the contribution of $\mathcal{C}=-1$ states, the model provides an excellent (and parameter-free) description of the lattice data with the Hagedorn temperature given by eq.~(\ref{Nambu_Goto_THagedorn}), which ensures consistency with the bosonic-string model. It is also worth noting that the lattice determination of the Hagedorn temperature following the method discussed in ref.~\cite{Meyer:2009tq} yields $\THagedorn \simeq 1.1 \Tc$ in the large-$N$ limit~\cite{Bringoltz:2005xx}. From these observations, one sees that statistical models with slightly different details can simultaneously mimic the actual thermodynamics to good accuracy: in some cases, the exclusion of some heavy states can be accounted for by a lower value for the Hagedorn temperature, and \emph{vice~versa}. In the present work, in which we modelled interactions between hadrons in terms of excluded-volume effects, for the $\SU(3)$ theory we observe that assuming the Hagedorn temperature defined by eq.~(\ref{Nambu_Goto_THagedorn}), the particle radii turn out to be compatible with zero: this is consistent with the findings obtained in ref.~\cite{Caselle:2015tza}. For the $\SU(2)$ theory there is no obvious justification for taking $\THagedorn \neq \Tc$; if one, nevertheless, tries to fit the lattice data using values $\THagedorn > \Tc$, one finds that the resulting curve for the point-like model is in slightly better agreement with lattice data at low temperatures, but not at intermediate and higher temperatures.

As for extensions of the present work, it would be instructive to perform the same analysis for observables sensitive to specific quantum numbers, in order to check the effect of repulsive interactions in full QCD and fit the size of hadronic states, as was suggested in ref.~\cite{Alba:2016hfo}. A qualitative comparison of excluded-volume-hadron-resonance-gas calculations to a set of lattice QCD observables was presented in ref.~\cite{Vovchenko:2016rkn}, but it would be interesting to implement flavor-dependent repulsions, as was done in ref.~\cite{Alba:2016hwx}: this work gave hints that such a physical picture is motivated by experimental data. We plan to address these issues in future work.

\vskip1.0cm 
\noindent{\bf Acknowledgements}\\
The lattice simulations were run on machines of the Istituto Nazionale di Fisica Nucleare (INFN) Pisa GRID Data Center and of the Consorzio Interuniversitario per il Calcolo Automatico dell'Italia Nord Orientale (CINECA). This research is supported by the Helmholtz International Center for the Facility for Antiproton and Ion Research (HIC for FAIR) within the Landes-Offensive zur Entwicklung Wissenschaftlich-\"okonomischer Exzellenz program of the State of Hesse. H.~S. acknowledges support from the Judah M. Eisenberg Laureatus Chair at Goethe University. We thank Michele~Caselle and Tamar~Friedmann for helpful discussions.

\bibliography{paper}

\end{document}